\documentclass[12pt]{article}

\usepackage{graphicx}
\usepackage{slashed}
\usepackage{amsmath,amssymb}
\usepackage{bm}
\usepackage{epsfig}

\newcommand{\bpsi}{\bar{\psi}}

\begin{document}

\title{
 SU(2)   Higher-order effective quark interactions from polarization 
}

\author{ F\'abio L. Braghin
% $^{1,2}$
\\
%\affiliation{$^1$
{\normalsize Instituto de F\'\i sica, Federal University of Goias,}
\\
{\normalsize Av. Esperan\c ca, s/n,
 74690-900, Goi\^ania, GO, Brazil}
%\\
%and 
%\\
%$^2$ Instituto de F\'\i sica, Univ. de S\~ao Paulo, Rua do Mat\~ao, Travessa R, 187,
%  05508-090, S\~ao Paulo, SP, Brazil
}

\maketitle

\begin{abstract}
%% Text of abstract
Higher order  quark effective interactions
are found for  SU(2) flavor by  departing  from a non local 
quark-quark interaction.
By integrating out a component of the  quark field, the determinant  is expanded in chirally symmetric 
and symmetry breaking
effective interactions up to  the  fifh order in the quark bilinears
 The resulting 
coupling constants are resolved in the leading
order of the   longwavelength limit 
and exact numerical ratios between several of these coupling constants
are obtained in the large quark mass limit.
%Whereas, in the present calculation, 
In  this level, chiral invariant interactions only 
show up in even powers of the quark bilinears, i.e.
${\cal O}(\bpsi \psi)^{2 n}$  ($n=1,2,3,..$), whereas
(explicit) chiral symmetry breaking terms emerge  as
${\cal O}(\bpsi \psi)^{n}$ being  always proportional to some power of 
the Lagrangian quark mass.
\end{abstract}

%\begin{keyword}
%Quark  Effective interaction \sep 
%Higher order interactions \sep
%Nambu Jona Lasinio model \sep
%Chiral symmetry \sep
%Explicit symmetry breaking 
%\end{keyword}

\section{Introduction}

The  
understanding of the effects and  mechanisms by which quarks
 interact among themselves 
 is a necessary step to provide a complete description of 
hadron structure and dynamics and  the phase diagram 
of Strong Interactions.
In low and intermediary energies these interactions can be  
parametrized  in terms of realistic  effective quark interactions
that  usually  
provide important information to establish
 the needed relations
between QCD and  
hadron dynamics
\cite{brambilla-etal,greensite}.
The basic and fundamental  mechanisms that give rise to 
each of the effective interactions and parameters present
in  
effective  models  and theories
should be expected to be  well
understood, although
a quite large amount of different quark effective interactions are expected to emerge
due to the intrincated structure of QCD.
The Nambu Jona Lasinio (NJL) model is known to describe 
qualitatively well  several important 
effects in hadron phenomenology \cite{NJL1,NJL}  in spite of its known 
 limitations.
A  large variety of possible corrections to the NJL coupling
can be expected to emerge from 
QCD, and higher order quark  interactions were shown to provide relevant effects
for the ground state 
\cite{8th,osipov,andrionov-andrionov},
 chiral phase transition
(flavor SU(2) and SU(3)) and higher energies
\cite{phtrans-su3,ho-su2,ho-su2-PL-eNJL,ho-su2-scd,osipov-ph-trans}
and eventually they might contribute to    multiquark structures \cite{tetraquarks}.
 In FAIR-GSI the high density phase diagram will be tested eventually 
providing relevant information also about the  role of multiquark interactions
in different regions of the phase diagram.
Few
mechanisms have been shown 
to drive quark  effective interactions by gluon exchange
\cite{thooft-det,creutz,NJL-derivations,kondo -2010,simonov-plb,qqq-instantons,simonov-prd,wetterich-etal,PRD-2014,vanderbossche,kashiwa-etal}.
 Instanton mediation have been shown to provide one
of the most investigated mechanisms for effective quark interactions
for example
by means of the Kobayashi-­Maskawa­-'t Hooft interaction or  instanton gas model.
It depends strongly on flavor and, for  flavor SU(2), it yields 
a second order quark interaction different from the usual chiral NJL interaction,
producing  the axial anomaly and its phenomenological consequences
 \cite{thooft-det,KKM,NJL1,NJL}.
%wtih correct phenomenological consequences \cite{thooft-anomaly}.
Polarization effects were shown to 
produce  low energy  and higher order
  effective interactions  \cite{PRD-2014}.

In the present work, flavor SU(2)   higher order
 quark effective interactions  
are calculated from  polarization effects 
by departing from 
a dressed one gluon exchange 
(i.e.  a global color model)
 along the lines of  Refs. \cite{PRD-2014}.
Simple gluon exchange is a basic mechanism that cannot describe low energy
hadron  properties, including dynamical breakdown of 
chiral symmetry (D$\chi$SB), although it can be dressed by
 gluon    interactions  producing  enough strength for 
 D$\chi$SB
 \cite{cornwall-2011,kondo,SD1}.
This work is organized as follows.
In the next section the method is shortly described
according to which the quark bilinears  are  separated into two 
components, 
i.e. $\bpsi \Gamma \psi \to (\bpsi \Gamma \psi)_1 + (\bpsi \Gamma \psi)_2$,
as done  in the background field method \cite{background}.
The background field ($\psi_1$)  remains as interacting 
quarks and the field $\psi_2$ is integrated out.
Instead of introducing auxilary fields (a.f.) for the component that is integrated out,
a weak field approximation is considered such that:
$(\bar{\psi} \psi)_1^2  >>  (\bar{\psi} \psi)_2^2$.
Results are the same as by introducing a.f.  in the leading order since the a.f.,
for example as shown  in Ref. \cite{PRD-2014,PRC,ERV},
 play  no role in the resulting leading
quark-quark effective   interactions.
The quark determinant is expanded in powers 
of quark bilinears yielding chiral invariant 
and also symmetry breaking terms proportional to the Lagrangian quark mass.
The corresponding effective couplings are resolved.
This expansion is performed up to the eighth  order  for all the bilinears 
 and up to the tenth order for the scalar-pseudoscalar ones.
Some ratios between the effective coupling constant 
are shown to provide simple numerical values.
Some numerical estimations are also shown.

\section{Diquark interaction and quark field splitting}

The departing point is  the following  
quark effective interaction:
%\cite{PRC,ERV}:
\begin{eqnarray} \label{Seff}  
S_{eff}  [\bar{\psi}, \psi] &=& 
\int_x \left[
\bar{\psi} \left( i \slashed{\partial} 
- m \right) \psi 
- 
 \frac{g^2}{2}\int_y j_{\mu}^b (x) 
{\tilde{R}}^{\mu \nu}_{bc} (x-y) j_{\nu}^{c} (y) 
%+  \frac{c\phi^2}{4}
 \right]
,
\end{eqnarray}
Where $b,c$ stand for color indices,
 the  color  quark current is 
$j^{\mu}_b = \bar{\psi} \lambda_b \gamma^{\mu} \psi$, 
 the sum in color, flavor and Dirac indices are implicit, $\int_x$ 
stands for 
$\int d^4 x$,
the kernel ${\tilde{R}}^{\mu \nu}_{bc}$ 
can be  written  in terms of 
transversal and longitudinal components 
($R_T$ and $R_L$) as:
$\tilde{R}^{\mu\nu}_{ab} \equiv \tilde{R}^{\mu\nu}_{ab} (x-y) = \delta_{ab} \left[
 R_T \left( g^{\mu\nu} - \frac{\partial^\mu \partial^\nu}{\partial^2}
\right) 
+ R_L \frac{\partial^\mu \partial^\nu}{\partial^2} \right]
$ 
 with  implicit Dirac delta functions $\delta(x-y)$.
With a Fierz transformation \cite{NJL1,NJL,PRC,ERV},
by picking up  the color singlet sector only,
the above effective quark interaction   can be expressed
in terms of 
 bilocal quark bilinears,
$j_i^q(x,y) =  \bar{\psi} (x) \Gamma^q \psi (y)$ where 
$q=s,p,v,a$
and  $\Gamma_q$ stands for Dirac and  flavor SU(2) 
operators
$\Gamma_{s} = I$ for the 2x2 flavor and 4x4  identities,
 $\Gamma_{p} = \sigma_i i  \gamma_5$,
$\Gamma_{v}^\mu =  \gamma^\mu \sigma_i $ and
$\Gamma_{a}^\mu =    i \gamma_5  \gamma^\mu  \sigma_i $,
being    $\sigma_i$ are the flavor SU(2) Pauli matrices.
The Fierz transformed interaction is written as:
$\Omega = \alpha \sum_q  j_i^q(x,y) R_q (x-y) j_i^q(y,x)$, where 
 $\alpha=8/9$, 
$R_q$ are the  kernels  in each of the $q$ channel of the interaction.
Next the quark field is separated into two components,
one of them   associated with polarization virtual processes
eventually to the formation of quark bound states such as light mesons 
and the chiral condensate and the other component remains as (constituent) quark. 
This procedure is basically  the one loop  background field method \cite{background},
and this will be done by rewritting the quark bilinears above as:
\begin{eqnarray} \label{split-Q} 
\bar{\psi} \Gamma^q \psi \to (\bar{\psi} \Gamma^q \psi)_2 
+ (\bar{\psi} \Gamma^q \psi)_1.
\end{eqnarray}
The Fierz transformed non local  interaction above can then be written as:
$\Omega \to \Omega_1 + \Omega_2 + \Omega_{12}$
where $\Omega_{1}$ and $\Omega_2$ stand for the interactions
of  each of the quark components, 
and
 $\Omega_{12}$ for the mixed terms.
The component $\psi_2$ will be integrated out 
and the fourth order terms can be eliminated in different approximated ways.
Firstly by simply considering a weak field approximation and therefore
by neglecting  $\Omega_2  << \Omega_1$. 
This yields the same results 
as the leading terms resulting from  the 
auxiliary field method which eliminates the fourth order interactions
$\Omega_2$,
as discussed in Refs. \cite{PRD-2014,PRC,ERV}.
In this case,  bilocal 
auxiliary fields ($S,P_i,V_\mu^i, \bar{A}_\mu^i$) are introduced
which   couple to the remaining 
 quark component. 
These couplings 
encode the non linearities of the 
initial model. 
However in this work we are interested only in the quark self interactions
and these couplings can be neglected. 
Even if one were interested in 
the effective interactions induced by these couplings to the auxiliary fields (a.f.),
the resulting 
quark-quark effective
interactions induced by the a.f. 
would be of higher order and  
numerically smaller.
By integrating out the 
 component $(\psi)_2$,
and by writing the determinant as:
 $\det (A) = \exp \left( Tr \ln A \right)$,
 the following non linear non local effective action 
for quarks $(\psi)_1$    is obtained:
%% By integrating out $\psi_2$,
\begin{eqnarray} \label{Seff-q4}  
S_{eff}  &=& - i \; Tr \ln \left\{ i
({S_0})^{-1} (x-y)
\right.
\nonumber
\\
&+& 
\left.
-
i \alpha g^2 \bar{R}^{\mu\nu} (x-y) \gamma_\mu  \sigma_i 
\left[
 (\bar{\psi}_y \gamma_\nu  \sigma_i \psi_x)
-  i \gamma_5   (\bar{\psi}_y
i \gamma_5 \gamma_\nu  \sigma_i \psi_x) 
\right]
\right.
\nonumber
\\
&+& 
\left.
 2 i  \alpha g^2 R(x-y)
 \left[  (\bar{\psi}_y \psi_x)
+ i  \gamma_5 \sigma_i  (\bar{\psi}_y i \gamma_5  \sigma_i \psi_x) \right]
 \right\}
- I_0
,
\end{eqnarray}
where
% the quark determinant, 
%$I_d= - i Tr \ln \left[ i {S}^{-1} \right]$,
  $Tr$ stands for traces of discrete internal quantum numbers indices
and integration of  spacetime coordinates/momentum
and 
\\
$I_0 = \int_x 
\left[
 \bar{\psi} \left( i \gamma \cdot \partial
- m \right)\psi
- \frac{g^2 }{2} \int_y j_{\mu}^{a}(x) R^{\mu\nu}_{ab} (x-y) j_{\nu}^{b}(y)
\right]$.
In this expression the label $_1$ for the quark field was omitted 
because  it is the only one remaining from here on.
$({S_0})^{-1} = ({S_0})^{-1}(x-y)  \equiv   (  i \gamma \cdot { \partial}
-  m )$, with an implicit Dirac delta function,
and where instead of $m$ one could introduce an effective mass ($m^*$)
which arise from the coupling to the scalar auxiliary variable $s$
 which produces
 the dynamical chiral symmetry breaking as discussed at length in Refs.
 \cite{NJL1,NJL,PRD-2014,PRC,ERV}.
The following kernels have also been defined from the Fierz transformation:
%\begin{eqnarray} \label{Rbar-Rbar}
$ R = R(x-y) = 
3 R_T  + R_L$ and 
$\bar{R}^{\mu\nu} =\bar{R}^{\mu\nu}(x-y) 
=  g^{\mu\nu} (R_T+R_L) + 
2 \frac{\partial^{\mu} \partial^{\nu}}{\partial^2} (R_T - R_L)$
with implicit Dirac delta functions.
By neglecting the derivative couplings,
with a shorthand notation for which the non local character of all the kernels is omitted, i.e
$R=R(x-y)$, $\bar{R}^{\mu\nu}=\bar{R}^{\mu\nu}(x-y)$ and $S_0=S_0(x-y)$,
 the quark determinant above can be rewritten \cite{mosel} as:
\begin{eqnarray} \label{Ideterm}
I_d &\equiv &
 -  \frac{i}{2}
Tr \ln \left[ S^{-1} {S^\dagger}^{-1} \right]
= - \frac{i}{2} Tr \ln [ \tilde{S}_0^{-1} ] 
\\
&-& \frac{i}{2} Tr \ln \left[1  + \beta   \tilde{S}_0
\left(  2   R \bpsi \psi 
- \bar{R}^{\mu\nu} \gamma_\mu \sigma_i \bpsi \gamma_\nu \sigma_i \psi \right)
+  g^4 \sum_{q,q'} \tilde{S}_0 a_{q,q'}
(\Gamma_q  \bpsi \Gamma_q \psi  ) 
( \Gamma_{q'}^\dagger
  \bpsi \Gamma_{q'} \psi  )
 \right]  
,
\nonumber
\end{eqnarray}
where 
$\beta= 2  m  g^2 \alpha$ was defined for the quark mass (symmetry breaking term),
$\tilde{S}_0 \equiv \tilde{S}_0 (x-y) =  - 1/ (\partial^2 + m^2) \delta (x-y)$ 
was factorized producing an irrelevant
multiplicative
constant  in the generating functional,
 $a_{q,q'}$ are  coefficients
for each of the flavor channels, 
 and crossed terms ($q,q'=s,p,v,a$)
with   the corresponding operators $\Gamma_q$ 
 and kernels $R_q$.
 This expression still has a strong non local character which is not written explicitely.
This determinant
will  be expanded for small $\tilde{S}_0$, i.e. large quark (effective) mass
by considering that $m$ may  be an effective (constituent) quark mass.
A small coupling $g^2$ or weak quark field $\psi_1$ 
yields  essentially the same results
 such that 
the final polynomial quark effective interactions are written in terms of effective coupling constants
in the local limit of the resulting couplings.
 It can be noticed that all the chiral invariant interactions
only appear from the contributions exclusively of the last term inside of the 
determinant.
 Therefore chiral invariant terms for this 
$SU(2)$ flavor will be $ {\cal O} [(\bpsi \psi)^2 ]^n$.
All the interactions for which the second term contributes
(proportional to 
the quark mass)  will  be not chiral invariant.
One of the 
 first order terms yields a  contribution for the  quark effective mass \cite{PRD-2014}
 of the form: $\Delta m^* = - i 2   \alpha g^2  m  \; Tr \; \tilde{S}_0 R$.

\section{ SU(2)  quark  effective  interactions }
\label{sec:eff-param}

The leading terms,
by resolving the effective coupling constants in the longwavelength limit
and 
 the zero order derivative expansion, are:
\begin{eqnarray} \label{4quark}
{\cal L}_{4} =
 g_4
\left[ ( \bar{\psi}  \psi )^2
 + (  \bpsi \sigma_i i \gamma_5 \psi )^2  \right]
-
 g_{v4} 
\left[ ( \bpsi  \sigma_i \gamma_\mu \psi )^2
 + ( \bpsi \sigma_i    \gamma_5 \gamma_\mu \psi )^2
\right]
+ {\cal L}_4^{sb}
\end{eqnarray}
where ${\cal L}_4^{sb} = g_{4,sb}
( \bar{\psi}  \psi )^2
+
g_{4,v,sb}  ( \bpsi  \sigma_i \gamma_\mu \psi )^2$
are  symmetry breaking terms which emerge from 
the second order expansion although they are  of the same order of magnitude as 
the first one, as it can be noted in the next expressions.
These effective coupling constants 
were resolved as:
\begin{eqnarray} \label{g4}
g_{4} \;  (1 \;  ; \; \delta_{i_j})  &=& 
- i 2 (g^2  {\alpha})^2  N_c\; Tr'' \;   {\tilde{S}_{0}}  R^2 \;
(1 \; ; \; \sigma_i \sigma_j ),
\\ \label{g4sb}
g_{4,sb}   &=& 
 i 4 (g^2 {\alpha})^2 N_c \; Tr'' \; {m}^2  ({\tilde{S}_{0}}  R)^2 ,
\\
g_{v4} \; \delta_{ij} g^{\mu\nu} 
&=&
- \frac{i}{2} (g^2  \alpha)^2 N_c
\; Tr'' \;     {\tilde{S}_{0}}  \bar{R}^{\mu\rho} 
   \bar{R}^{\nu\sigma}
     (\sigma_i \sigma_j)
\gamma_\rho \gamma_\sigma 
\\ \label{g4vsb}
g_{4,v,sb} \; \delta_{ij} g^{\mu\nu} 
&=&
i (g^2   \alpha)^2 N_c
\; Tr'' \;   {m}^2  {\tilde{S}_{0}}  \bar{R}^{\mu\rho} 
 {\tilde{S}_{0}}  \bar{R}^{\nu\sigma}
     (\sigma_i \sigma_j)
\gamma_\rho \gamma_\sigma ,
\end{eqnarray}
where 
 where $Tr''$  includes  all the traces in internal and spacetime indices except 
 the trace in color indices that has  already been done.
The couplings with $g_4$ and $g_{v4}$  are 
the usual  NJL and vector NJL couplings respectively 
with  dimension $1/M^2$ for a mass scale $M$.
For the class of diagrams of this one fermion loop level, 
by considering that 
 $g^2 \sim \tilde{g}^2/N_c$,
the resulting  $n$-quark coupling constants 
 are 
 of the order of $N_c^{1-n}$ in agreement with 
\cite{witten}.

The non derivative sixth order terms, after resolving the effective coupling
constants, are all symmetry breaking and they were found to be: 
\begin{eqnarray} \label{6th-1}
{\cal L}^{(6)}
&=&
 g_{6,sb}^{(1)}  (\bar{\psi} \psi )
\left[  (\bar{\psi} \psi )^2
  +   (\bar{\psi} \sigma_i i \gamma_5 \psi)^2  \right]
-   g_{6,sb,a}  \epsilon_{ijk}
 (\bar{\psi} \sigma_i i\gamma_5 \gamma_\mu \psi) (\bar{\psi} \sigma_j   \gamma^\mu \psi)
( \bar{\psi} \sigma_k i \gamma_5 \psi )
\nonumber
\\
&-&
g_{6,sb,a}  
\left[  
 ( \bar{\psi} \sigma_i \gamma_\mu \psi )^2  +  
( \bar{\psi} \sigma_i i \gamma_5 \gamma_\mu \psi )^2 
 \right]
 (\bar{\psi}  \psi )
+ g_{6,sb}^{(3)} (\bar{\psi} \psi)^3
\end{eqnarray}
where 
\begin{eqnarray}  \label{g6-all}
g_{6,sb}^{(1)}     \; (1 \; ; \;  \delta_{ij}) &=& 
 i  2  (\alpha g^2)^3 N_c
 \; Tr'' \; {m} {\tilde{S}_{0}}  \; R  ({\tilde{S}_{0}}  \; R^2)  \;
(1 \; ; \; \gamma^2_5  \sigma_i \sigma_j)
,
\\
g_{6,sb}^{(3)}     \; (1 \; ; \;  \delta_{ij}) &=&  - 
 i  \frac{32}{3}  (\alpha g^2)^3 N_c
 \; Tr'' \; {m}^3  ({\tilde{S}_{0}}  \; R)^3 
\; (1 \; ; \;  \sigma_{i} \sigma_j)
,
\nonumber
\\
g_{6,sb,a}  \; g^{\nu\sigma} 
( \delta_{ij} \; ; \;  i \epsilon_{ijk} ) 
 \; 
&=& 
  i ({\alpha} g^2)^3 N_c \; 
Tr'' \;  {m} {\tilde{S}_{0}}  R  {\tilde{S}_{0}}   R^{\mu\nu} 
 R^{\rho \sigma}\;
\gamma_\mu \gamma_\rho \gamma_5^2 
 \sigma_i \sigma_j \left( 1 \; ; \; \sigma_k \right),
\nonumber
\end{eqnarray}
where for further calculation one defines
$\bar{R}_2^{\nu\sigma} = \bar{R}^{\mu\nu} \bar{R}_{\mu}^{\sigma}
= (R_T +  R_L)^2  g^{\nu\sigma} + 8 R_T (R_T - R_L) 
\frac{\partial^\nu \partial^\sigma}{\partial^2}$.

There are several chiral invariant and 
 symmetry breaking 
 non derivative eighth order interactions.
They  were found to be:
\begin{eqnarray} \label{8th-1}
{\cal L}^{(8)}
&=&
g_8  
\left[  
(\bpsi \psi)^2
 +   (\bpsi i \gamma_5 \sigma_i \psi)^2  \right]^2
+
g_{8,sb}^{(2)}  ( \bar{\psi}  \psi )^2 
  \left[
 ( \bar{\psi}  \psi )^2
 + (  \bpsi \sigma_i i \gamma_5 \psi )^2  
   \right]
+ g_{8,sb}^{(4)}  ( \bar{\psi}  \psi )^4
\nonumber
\\
&+&
g_{8v}
\left[  
(\bpsi \gamma_\mu \sigma_i \psi)^2
 +   (\bpsi  \gamma_5  \gamma_\mu  \sigma_i \psi)^2 
\right]^2
-  g_{8v,sb}
(\bpsi \psi)^2
\left[ 
(\bpsi \gamma_\mu \sigma_i \psi)^2
 +   (\bpsi  \gamma_5  \gamma_\mu  \sigma_i \psi)^2 
\right]
\nonumber
\\
&-& g_{8vs}  \left[
(\bpsi \gamma_\mu \sigma_i \psi)^2
 +   (\bpsi  \gamma_5  \gamma_\mu  \sigma_i \psi)^2 
\right]
 \left[  (\bpsi \psi)^2
 +   (\bpsi i \gamma_5 \sigma_i \psi)^2 
\right]
\nonumber
\\
&+&
 g_{8}^{s} 
(\bpsi  \gamma_\nu \sigma_j \psi)^2 \left[
(\bpsi \gamma_\mu \sigma_i \psi)^2
 +   (\bpsi  \gamma_5  \gamma_\mu  \sigma_i \psi)^2 
\right]
,
\end{eqnarray}
%where  $E^{ijkl} = \delta_{ij} \delta_{kl} - \delta_{il} \delta_{jk}$.
where the chiral invariant terms are  of second order of the expansion, 
and the symmetry breaking are 
of third and fourth orders in the expansion 
of $I_d$.
Up to  this order of the expansion, 
terms in odd powers of the pseudoscalar and axial bilinears 
naturally disappear due to the traces such as
$tr (\gamma_5) = 0$.
The effective coupling constants are the following:
\begin{eqnarray}  \label{g8}
g_{8}\; (1 \; ;  \delta_{ij} ) &=& 
4  i   (\alpha g^2)^4 N_c 
 \; Tr'' \; ({\tilde{S}_{0}} \; R^2)^2
(1 \; ; \; \gamma^2_5  \sigma_i \sigma_j),
\nonumber
\\
g_{8,sb}^{(2)} (1\; ; \;\delta_{ij} ) &=& 
128 i   (\alpha g^2)^4 N_c
 \; Tr'' \; m^2  (\tilde{S}_0 R)^2 ({\tilde{S}_{0}} \; R^2)
(1 \; ; \; \gamma^2_5  \sigma_i \sigma_j) ,
\nonumber
\\
g_{8,sb}^{(4)} &=& 
64  i   (\alpha g^2)^4 N_c
 \; Tr'' \; m^4  ({\tilde{S}_{0}} \; R)^4 ,
\nonumber
\\
g_{8v}  \Gamma^{\mu_1\nu_1\rho_1\sigma_1}
 \Gamma_{ijkl}
&=& \frac{i}{2}
(\alpha g^2)^4 N_c \; Tr'' \; \tilde{S}_0 \bar{R}^{\mu_1\mu_2} \bar{R}^{\nu_1\nu_2}
\tilde{S}_0
 \bar{R}^{\rho_1\rho_2} \bar{R}^{\sigma_1\sigma_2}
(\gamma_{\mu_2}\gamma_{\nu_2} \gamma_{\rho_2} \gamma_{\sigma_2})
( \sigma_i \sigma_j \sigma_k \sigma_l),
\nonumber
\\
g_{8vs} g^{\mu\rho} \delta_{ij}
&=& -
i 4 (\alpha g^2)^4 N_c  \; Tr'' \; (\tilde{S}_0 R^2) (\tilde{S}_0 R^{\mu\nu} 
  R^{\rho\sigma} )
\sigma_i \sigma_j \gamma_\nu \gamma_\sigma,
\nonumber
\\
g_{8v,sb} g^{\mu\rho} \delta_{ij}
&=&
 - i 8 (\alpha g^2)^4  N_c \; Tr'' \; m^2 (\tilde{S}_0 R)^2 (\tilde{S}_0 R^{\mu\nu} 
  R^{\rho\sigma} )
\sigma_i \sigma_j \gamma_\nu \gamma_\sigma,
\\
g_{8}^{(s)}
 \Gamma^{\mu_1\rho_1}_{\mu\rho}
 \delta_{ijkl}
&=&
- \frac{i}{2}
 (\alpha g^2)^4  N_c \; Tr'' \; m^2  (\tilde{S}_0 R^{\mu_1\nu_1})
(\tilde{S}_0 R^{\rho_1\sigma_1})
 (\tilde{S}_0 R_{\mu}^{\nu} 
  R_{\rho}^{\sigma} )
\gamma_{\nu_1} \gamma_{\sigma_1}
\gamma_\nu \gamma_\sigma
(\sigma_i \sigma_j  \sigma_k \sigma_l),
\nonumber
\end{eqnarray}
where $\Gamma_{ijkl}= \delta_{ij}\delta_{kl} +   \delta_{il}\delta_{jk} 
-  \delta_{ik}\delta_{jl}$
and   
$\Gamma_{\mu\nu\rho\sigma}= g_{\mu\nu}g_{\rho\sigma} +   g_{\mu\sigma}g_{\nu\rho} 
+  g_{\mu\rho}g_{\nu\sigma}$.
Some of these terms were considered  
 in  Ref. \cite{ho-su2}.

The tenth order interaction terms
(leading terms from expansion up to the fifth order)
 are all symmetry breaking and
 the scalar-pseudoscalar terms can be written as:
\begin{eqnarray} \label{10th-1}
{\cal L}^{(10)}
=
g_{10}^{(1)}    
 ( \bar{\psi}  \psi ) \left[  
 ( \bar{\psi}  \psi )^2
 + (  \bpsi \sigma_i i \gamma_5 \psi )^2  \right]^2
+
g_{10}^{(3)}  ( \bar{\psi}  \psi )^3   \left[ ( \bar{\psi}  \psi )^2 
 + (  \bpsi \sigma_i i \gamma_5 \psi )^2     \right]
+ g_{10}^{(5)}  ( \bar{\psi}  \psi )^5 ,
\end{eqnarray}
where:
\begin{eqnarray} \label{g10}
g_{10}^{(1)}  &=&
-  \frac{i}{2}   (4\alpha g^2)^5 N_c
 \; Tr'' \; m ({\tilde{S}_{0}}  \; R) ( \tilde{S}_0 R^2 )^2 ,
\nonumber
\\
g_{10}^{(3)}  &=&
  \frac{i 3}{4} (4 \alpha g^2)^5 N_c
 \; Tr'' \; {m}^3  ({\tilde{S}_{0}}  \; R)^3 
 ({\tilde{S}_{0}}  \; R^2) ,
\nonumber
\\
g_{10}^{(5)}  &=&
- \frac{i}{10}  (4\alpha g^2)^5 N_c
 \; Tr'' \; {m}^5  ({\tilde{S}_{0}}  \; R)^5 ,
\end{eqnarray}
The symmetry breaking terms of the scalar-pseudoscalar channel
can be written 
 in a general form for the $n-$term of the expansion 
in terms of a  number  (combinatorial) $a_m$,:
\begin{eqnarray}
g_{2n,sb}^{(m)}  \; = \;
 \frac{i}{n}  a_m   (2\alpha g^2)^n
 \; Tr \; {m}^m ({\tilde{S}_{0}}^{\frac{(n+m)}{2}}  \; R^{n})
.
\end{eqnarray}

 One can consider two particular limits for calculating 
  ratios of the  quark effective coupling constants
depending on the   gluon propagator components. 
These ratios are obtained by  assuming a  large quark mass 
%such that
%$\tilde{S}_0 \sim - 1/m^2$ 
and by choosing one of the two  following 
limits:
(I) $R_L=0$ ($^T$),
or (II) $R_T=0$ ($^L$).
With the expressions shown above which turns out to
depend on the vector or axial bilinears,
 the moduli of some ratios yield:
\begin{eqnarray} \label{ratio}
  &&
\left| \frac{g_{4}}{g_{v4}} \right|^T 
\sim \left| 4 \frac{g_{4}}{ g_{4,v,sb}} \right|^T
\sim 3,
\hspace{.8cm} 
\left|
\frac{g_{6,sb}^{(1)}}{g_{6,sb,a}}
\right|^T
\sim 6 ,
\hspace{.8cm}
\left|
\frac{g_{8}}{g_{8vs}}
\right|^T
\sim  \frac{3}{4} ,
\\
&& 
\left|\frac{g_{4}}{g_{v4}}\right|^L 
\sim \left|4  \frac{g_{4}}{g_{4,v,sb}} \right|^L
\sim  1 ,
\hspace{.8cm}
\left|
\frac{g_{6,sb}^{(1)}}{g_{6,sb,a}}\right|^L
\sim  2 ,
\hspace{.8cm}
\left|\frac{g_{8}}{g_{8vs}}
\right|^L
\sim  \frac{1}{4} .
\end{eqnarray}
The ratios between the chiral invariant fourth order coupling constants 
$(\frac{g_{4}}{g_{v4}})$ are  in good agreement with phenomenology
 \cite{vector-NJL,sugano,deborah}.
These ratios might therefore present
 quite strong gauge dependence and this issue will not be discussed 
in the present work.
Some ratios are independent of the gluon kernel component
and their moduli are given by:
\begin{eqnarray} \label{ratio-2}
\left| \frac{g_4}{g_{4,sb}} \right| \sim \frac{1}{2},
\hspace{0.3cm} 
\left|\frac{g_{6,sb}^{(1)}}{g_{6,sb}^{(3)}}\right|
\sim \frac{3}{4} ,
\hspace{.3cm}
\left|\frac{g_{8}}{g_{8sb}^{(2)}}\right|
\sim 
\left|\frac{g_{8}}{2 g_{8sb}^{(4)}}\right|
\sim  \frac{1}{32},
\hspace{0.3cm} 
\left|\frac{g_{10}^{(1)}}{g_{10}^{(3)}}\right|
\sim  1,
\hspace{0.3cm} 
\left|\frac{g_{10}^{(1)}}{g_{10}^{(5)}}\right|
\sim   5,
\end{eqnarray}
the first of this ratios shows that the exclusive contribution of the 
explicit chiral symmetry breaking via the Lagrangian quark mass
for the coupling $(\bpsi \psi)^2$ is of the same order  of magnitude 
as  the NJL coupling.
Next, some numerical values are shown  by replacing the
traces in spacetime coordinates by momentum integration rotated to 
Euclidean space in the limit of zero momentum exchange.
A simplified confining gluon propagator from Ref. \cite{cornwall-2011} 
is considered 
with the same values for the prescription given by expression (10) of 
Ref. \cite{cornwall-2011}.
The only ultraviolet  divergent effective parameter  presented above
  is the one 
for the effective mass correction before Section (\ref{sec:eff-param}).
It can be directly renormalized 
with the Lagrangian mass counterterms and it will not be estimated here.
The mass for the quark kernel $\tilde{S}_0$
 was considered to be an effective mass from D$\chi$SB $m = 0.33$ GeV and
the
coupling constant $g^2$  as  the zero momentum limit of the 
QCD lattice calculations divided by 1000,
 i.e. $g^2  = 17.8 \pi/(10^3 N_c)$ from Ref. \cite{lattice}.
It is reasonable to consider a reduced value because a full 
running coupling constant
would reduce the contribution of the higher energy modes.
The resulting values 
were found to be $g_4 \simeq 1.2 $ GeV$^{-2}$, $g_6 \simeq - 28.2$ GeV$^{-5}$,
$g_8 \simeq 4.1 \cdot 10^{4}$ GeV$^{-8}$
and $g_{10}^{(1)} \simeq 2.2  \cdot 10^{8}$ GeV$^{-11}$.
These values are comparable to values 
obtained in the literature by phenomenological fitting
except the higher order ones.
 From  Ref. 
\cite{ho-su2} some SU(2) flavor coupling constants were considered as:
$g_4 \sim 10 $ GeV$^{-2}$ and $g_8 \sim 100 - 450$ GeV$^{-8}$, 
and for the sake of comparison for SU(3)
Refs. \cite{osipov}
 $g_4 \sim 10$ GeV$^{-2}$,  
$g_6 \sim - 1100$  GeV$^{-5}$,
 $g_8 \sim 6000$ GeV$^{-8}$. 
The values for the higher order couplings are somewhat larger than 
the values obtained from phenomenology and this might be related to the
 truncated momentum dependence considered
 and to the values of the parameters $m, g^2$ considered above.

The  emerging quark-quark potential 
is therefore composed by several types of chiral invariant and symmetry breaking
terms 
and this intrincated structure is expected from a confining theory
\cite{greensite}. 
Obvious corrections to the effective interactions found above
are due to the derivative interactions that were not calculated
and which may be expected to be relevant for a complete effective theory
for quark dynamics.
 It is interesting to emphasize two points: 
firstly it can be seen in expressions (\ref{g4sb},\ref{g4vsb},\ref{g6-all})  and 
the symmetry breaking couplings of expressions (\ref{g8}) and (\ref{g10}),
that all the symmetry breaking effective interactions
have the effective couplings proportional to the Lagrangian quark mass, that is the explicit symmetry breaking 
term. If the quark mass were corrected by the quark condensate to an effective quark mass
the same conclusion holds.
Secondly, the strength of the resulting symmetry breaking effective couplings are of the order of 
the chiral invariant terms.
The  expressions for  these effective quark 
 interactions were obtained 
without an explicit form of the gluon propagator
which  plays  a fundamental  role in the resulting relative strength of the 
resulting effective
coupling constants.
Furthermore all the expressions for the  effective coupling constants were written 
in a way to make possible to compute the corresponding form factors.
It is also interesting to emphasize that  results of this work allows for systematic computation of 
effective coupling constants without performing extensive phenomenological fits 
with hadron masses and/or couplings.
Although the gluon propagator and higher order gluon
 interactions in the departing quark effective action 
might induce different quark-quark effective interactions
they should  not be expected to change the shape of the effective interactions
found in the present work.

\subsection*{ Acknowledgement}

 F.L.B. thanks short conversation with G.I. Krein.
This work was
 partially supported by CNPq-Brazil 
(482080/2013-2).

\end{document}